\RequirePackage{fix-cm}
\documentclass[twocolumn]{svjour3}       
\smartqed  
\usepackage{graphicx}
\usepackage{float}
\usepackage{mathptmx}      
\usepackage{amsmath}
\usepackage{cite}
\usepackage{hyperref}
\hypersetup{colorlinks, linkcolor={black}, citecolor={black}, urlcolor={black}}
\newcommand{\pos}{\vec{x}}
\newcommand{\vel}{\vec{v}}
\journalname{Computational Particle Mechanics}
\begin{document}

\title{Particle-based simulation of ellipse-shaped particle aggregation as a model for vascular network formation\thanks{The investigations were in part supported by the Division for Earth and Life Sciences (ALW) with financial aid from the Netherlands Organization for Scientific Research (NWO) and by The Netherlands Consortium for Systems Biology (NCSB), which is a part of the Netherlands Genomics Initiative/Netherlands Organisation for Scientific Research. AS acknowledges support by Marie Curie FP7 IEF (grant number 329968). The authors acknowledge the use of the UCL Legion High Performance Computing Facility (Legion@UCL) in the completion of this work.}}


\author{Dimitrios Palachanis \and
        Andr\'{a}s Szab\'{o} \and
        Roeland M.H. Merks
}


\institute{Dimitrios Palachanis \at
			  Centrum Wiskunde \& Informatica, 
			  Science Park 123, 
			  1098 XG Amsterdam, The Netherlands\\
			  Leiden Institute of Advanced Computer Science, Leiden University, Niels Bohrweg 1, 2333 CA Leiden, The Netherlands \\
              \emph{Present address:} Aristotle University of Thessaloniki,
              Faculty of Sciences, 
              Department of Chemistry,
              University Campus, 54124, Thessaloniki, Greece\\
              Tel.: +30-2310-999206\\ 
              \email{d.palachanis@gmail.com} \\
            \and
            Andr\'{a}s Szab\'{o} \at
			  Centrum Wiskunde \& Informatica, 
			  Science Park 123, 
			  1098 XG Amsterdam, The Netherlands\\
              \emph{Present address:} Department of Cell and Developmental Biology,
              University College London, 
              London, UK\\
              \email{A.Szabo@ucl.ac.uk}
           \and
           Roeland M.H. Merks \at
              Centrum Wiskunde \& Informatica, 
              Science Park 123, 
              1098 XG Amsterdam, The Netherlands\\
              Mathematical Institute, Leiden University, Niels Bohrweg 1, 2333 CA Leiden, The Netherlands \\
              \email{roeland.merks@cwi.nl} \\
}

\date{Received: date / Accepted: date}
\maketitle

\begin{abstract}
Computational modelling is helpful for elucidating the cellular mechanisms driving biological morphogenesis. Previous simulation studies of blood vessel growth based on the Cellular Potts model (CPM) proposed that elongated, adhesive or mutually attractive endothelial cells suffice for the formation of blood vessel sprouts and vascular networks. Because each mathematical representation of a model introduces potential artifacts, it is important that model results are reproduced using alternative modelling paradigms. Here, we present a lattice-free, particle-based simulation of the cell elongation model of vasculogenesis. The new, particle-based simulations confirm the results obtained from the previous Cellular Potts simulations. Furthermore, our current findings suggest that the emergence of order is possible with the application of a high enough attractive force or, alternatively, a longer attraction radius. The methodology will be applicable to a range of problems in morphogenesis and noisy particle aggregation in which cell shape is a key determining factor.
\end{abstract}

\section{Introduction}

The vascular system is one of the most important organs in large multicellular organisms. During embryogenesis the need for an efficient transport of nutrients and waste products arises naturally as the increasing size of the organism makes diffusion less and less efficient. A similar situation arises in solid tumours, but instead of the de novo formation of the vasculature (vasculogenesis), tumours are able to remodel the vascular network in the surrounding tissues and promote formation of new branches from the existing ones (angiogenesis)~\cite{hanahan_hallmarks_2011}. Understanding the basic principles behind these processes could help in controlling tumour growth, as well as improving restoration of normal vasculature after trauma such as a stroke. It is tempting to speculate that the same basic principle that generally promotes network formation during vasculogenesis also drives blood vessel sprouting during angiogenesis.

To explore mechanisms behind vasculogenesis it is beneficial to turn to computational modelling where all parameters are under control, before attempting experimental validation where unknown factors could complicate the interpretation of the results. Various hypotheses have been constructed with the aid of computational biology~\cite{czirok_endothelial_2013}, however, we argue that the best modelling approach to study network formation is through cell-based models~\cite{merks_cell-centered_2005}. Such models are well suited to describe collective behaviours emerging from cellular properties and interactions, such as collective cell motion~\cite{mehes_collective_2014}, or various morphogenetic processes~\cite{merks_modeling_2009}. Vascular networks contain structures typically on the scale of a couple to tens of cells which would invalidate continuum-based approaches, that are better fitted for describing cellular structures of at least hundreds of cells in size. 

Previous studies using cell-based modelling have identified a handful of candidate mechanisms for sprouting and network formation~\cite{czirok_pattern_2012}. By the nature of biological cells, all of these mechanisms include a form of long-range attraction that ensures cell cohesion, and a short-range repulsion that is needed to represent cell incompressibility and thus prevent unrealistically high cell densities. One class of models explains the long-range attraction via mechanotaxis, whereby cells contract and deform an underlying soft substrate, pulling cells towards each other \cite{Manoussaki1996} or where the substrate locally stiffens in response to cellular contraction directing cells towards one another via durotaxis~\cite{VanOers2014}. This mechanism suffices to explain how in some culture systems, for example bovine aortic endothelial cells on poly-acrylamide gels~\cite{VanOers2014}, networks form only on substrates of specific stiffness. Contractile cells on deformable substrates have been described as force dipoles revealing that such dipoles tend to form locally or globally ordered structures depending on the Poisson ratio of the substrate \cite{Bischofs2005, Bischofs2006}. An alternative mechanism, based on the observation that cells preferentially adhere to elongated cellular structures, explains how networks can form on bare substrates~\cite{szabo_network_2007,szabo2008mathematical}. Another body of work focuses on chemotaxis towards a self-secreted chemokine, in combination with either contact-inhibition of chemotaxis \cite{merks_contact-inhibited_2008} or cell elongation \cite{merks_cell_2006}, that proves to be a robust mechanism albeit the chemokine remains to be identified. While the different mechanisms of attraction distinguish between different hypotheses, the mechanism of repulsion is less important and is either introduced explicitly (for example in partial-differential equation models~\cite{Manoussaki1996} or particle-based models~\cite{szabo_network_2007}) or is an implicit property of the model (as in the CPM \cite{VanOers2014,merks_cell_2006} for instance).
\begin{figure}
\centering\includegraphics[width=\columnwidth]{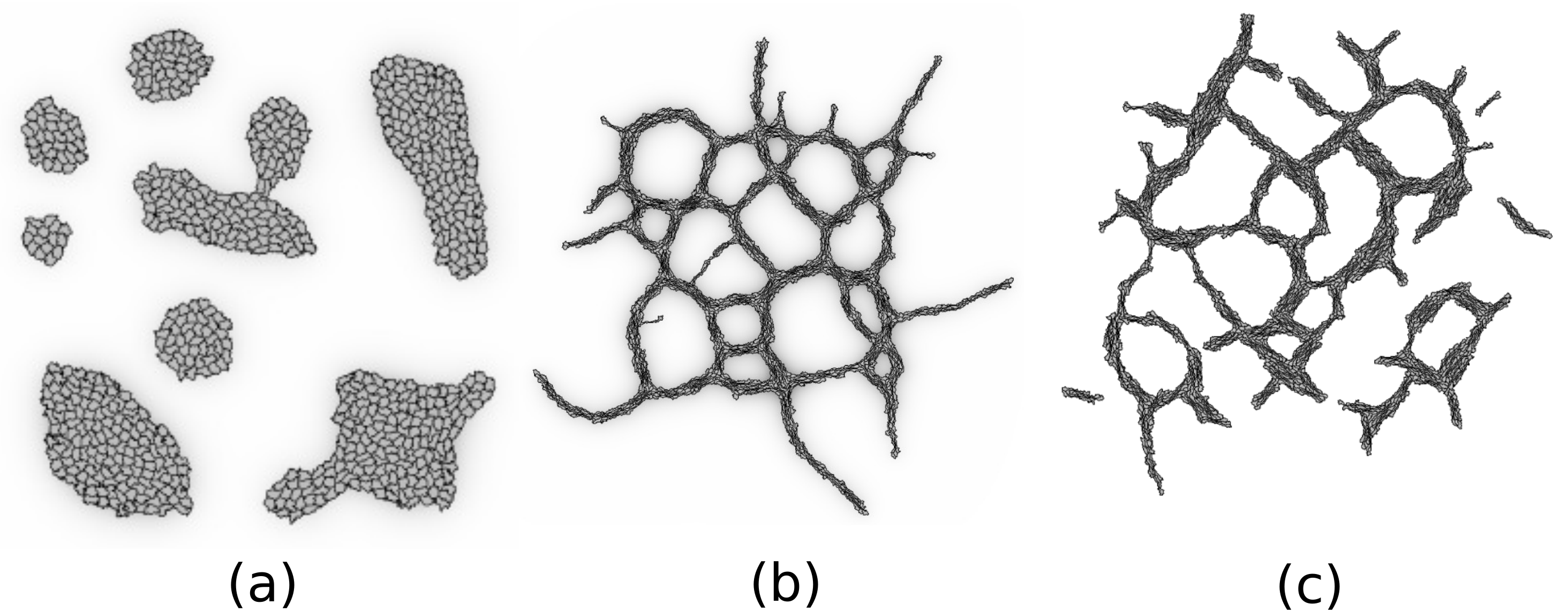}
\caption{\textbf{Cellular Potts simulations of chemotaxis and cell elongation models of vasculogenesis.} 
\textbf{a}, A simple mechanism of chemotaxis towards a self-secreted chemokine generates cell aggregates. 
\textbf{b}, Chemotaxis together with cell elongation gives rise to robust networks.
\textbf{c}, Cell elongation without chemotaxis is sufficient to explain network formation. 
Used with permission from \cite{palm_vascular_2013}}
\label{fig:palm_merks}
\end{figure}

The model mechanism we focus on in this paper demonstrates that cell elongation and volume exclusion together with contact-dependent cell adhesion~\cite{palm_vascular_2013} or longer range cell--cell attraction~\cite{merks_cell_2006}, are sufficient for network formation (Fig.~\ref{fig:palm_merks}). Whereas slightly adherent, dispersed cells aggregate into compact clusters (Fig.~\ref{fig:palm_merks}a), elongated cells form network-like structures in the models with (Fig.~\ref{fig:palm_merks}b) or without (Fig.~\ref{fig:palm_merks}c) additional longer range chemotaxis. In these models of elongated cells, the branches are stabilised by the increasing rotational inertia of growing clusters of elongating cells: a tightly packed branch of elongated cells is harder to displace or reorganise, therefore cells are frozen into the branched pattern in the model. A somewhat similar cellular process to elongation described by this model is the apical constriction of epithelial cells that plays a major role in, for example, gastrulation and neurulation~\cite{sawyer_apical_2010}. Although apical constriction leads to cell elongation, this elongation may only be a passive result of volume conservation and cytoplasmic flow~\cite{he_apical_2014}.

As each model implementation is a simplification with different limitations, it is worth to examine the model at hand using more than one implementation with different implicit assumptions associated with the simulation methodology. If the hypothesis is independent of the implementation, it should be able to reproduce the phenomenon in different implementations. For example, this has been performed in the case of the preferential adhesion hypothesis, which has been tested in both a lattice-free~\cite{szabo_network_2007} and a grid-based model~\cite{szabo2008mathematical}. 

The previous studies \cite{merks_cell_2006, palm_vascular_2013} used the same cellular Potts model (CPM) description which could introduce model-specific artefacts. For example, highly elongated cells could breach the limitations of cell representation within the CPM, as cell elongation is achieved by maximizing the largest moments of inertia of the cell\cite{merks_cell_2006}, which leads to unrealistic thickening of the cell body at its extremities. As artefacts may result from the grid-based nature of the model and the implementation of cell elongation, we chose to implement the cell elongation hypothesis in a lattice-free model using a generalized attraction and repulsion between cells, describing a class of models. We use this model to test if the hypothesis holds and to compare our results to previous reports.

\section{Methods}
\subsection{Computational Model}

\begin{figure}
\centering\includegraphics[width=\columnwidth]{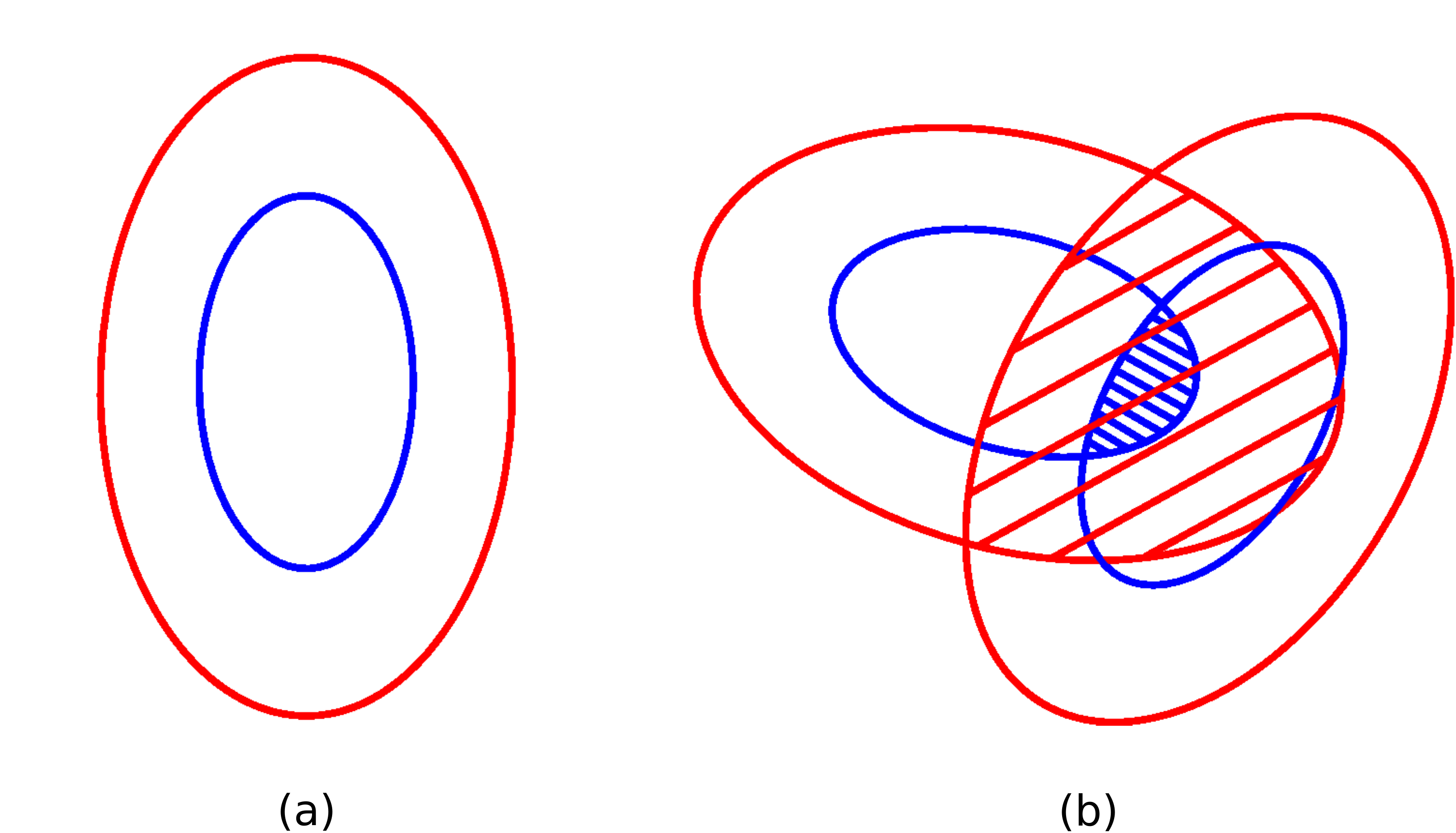}
\caption{\textbf{Cells and their interaction in the model.} 
\textbf{(a)} Cells are represented as 2D ellipses consisting of a repulsive core as cell body (blue). Cell attraction occurs at a finite range around the core in an area represented as a bigger ellipse surrounding concentrically the repulsive core (red).
\textbf{(b)} Cell-cell interaction is governed by the overlap area of the two inner ellipses (blue area) and the overlap of the two outer ellipses (red area)}
\label{fig:ell_ell_overlap}
\end{figure}

Each cell $i$ in the model is described as an ellipse on a 2D plane, characterized by its position $\pos_i$, direction of elongation $\phi_i$, area $A_i$, and the aspect ratio of major-to-minor axis $s_i$. This ellipse is a repulsive core representing the incompressible cell body (Fig. \ref{fig:ell_ell_overlap}a, blue) and is surrounded by a larger concentric ellipse that is responsible for cell-cell attraction in the model (for example via filopodial adhesion) (Fig. \ref{fig:ell_ell_overlap}a, red). Cell motion is modelled as a persistent diffusion process as in a previous study~\cite{szabo_network_2007}. The change in velocity $\vel_i(t)$ for cell $i$ at time $t$ is described as:
\begin{equation}
\label{eq:diff1}
\frac{d\vel_i(t)}{dt}=\frac{1}{m_i}\left(-\alpha\vel_i(t)+N_v\vec{\xi}_v dt^{-0.5}+\sum_{j\neq i}^{j} \frac{\pos_i-\pos_j}{|\pos_i-\pos_j|} F_{ij}\right)
\end{equation}
Here $m_i$ is the mass of cell $i$ ($m_i=m=1$), $\alpha$ is a damping parameter. $\vec{\xi}_v$ is a uniformly distributed random vector in the 2D plane ($\vec{\xi}_v \in [-0.5:0.5] \times [-0.5:0.5]$), and model parameter $N_v$ sets the amplitude of this translational noise. The last term is the sum of pair interactions between cells describing short-range core repulsion, and a long-range but finite attraction:
\begin{equation}
\label{eq:interaction}
F_{ij}=\lambda_r A_{r} - \lambda_a A_{a}.
\end{equation}
$A_r$ is the overlap area between the two repulsive ellipses of the two cells (Fig.~\ref{fig:ell_ell_overlap}b, blue area), and $A_{a}$ is the overlap area of the two outer ellipses (Fig.~\ref{fig:ell_ell_overlap}b, red area), and $\lambda_r$ and $\lambda_a$ are model parameters. The overlap is calculated using a previously published method \cite{ee_overlap}.

Model cells are rotated to minimize overcrowding and maximize attraction \cite{Bischofs2006}. For every interval $\Delta t$ an attempt is made to change the orientation of N randomly selected cells with a random angle $\xi_a\sqrt{\Delta t}$, where $\xi_a \in [-\pi:\pi]$ is a uniform random variable. The change is accepted with a turn-probability
\begin{equation}
\label{eq:rotationProb}
P_i= \min \left[1,\exp \left\{\frac{1}{N_a}\left(\sum_{j\neq i}F_{ij} - \sum_{j\neq i}F_{ij}'\right)\right\}\right],
\end{equation}
where $F_{ij}'$ is the pair interaction in a configuration after the proposed rotation and $N_a$ is the angular noise parameter. 

Simulations were initiated with $N=1000$ overlapping cells distributed at random within a $200\;\mathrm{\mu m}\times 200\;\mathrm{\mu m}$ unit area. Cell area was fixed at $250\;\mathrm{\mu m^2}$ throughout the study. The size of the outer ``adhesive" ellipse is set by the attraction radius parameter defined as $R_a = a_{a} / a$, where $a_a$ is the semi-major axis of the adhesive (outer) ellipse and $a$ is the semi-major axis of the repulsive (inner) ellipse, and the aspect ratio of the outer and inner ellipses is the same. Baseline parameters used are: $R_a=1.5$, $\lambda_a=0.0006,  \lambda_r=0.02, \alpha=0.4, N_v=0.4$, $N_a=1$ and $s=7$. These parameters were estimated by macroscopic inspection of the model to yield ``realistic"  pattern formation as a measure of validation. To prevent excessive cell overlap, the magnitude of the repulsion strength ($\lambda_r$) is two orders of magnitude larger than the magnitude of the attraction strength ($\lambda_a$). The model's sensitivity to the parameters was tested by altering the parameter values until unrealistic results, either concerning cell movement or cell interactions, were produced. Model equations \ref{eq:diff1} and \ref{eq:interaction} are integrated numerically using the forward Euler method with a fixed time-step of $\Delta t=0.1$ (or $\Delta t=1$ for the longer simulations of up to $t=100,000$ shown in Fig.~\ref{fig:temporal_order}), and orientations are updated after each iteration synchronously using Eq.~\ref{eq:rotationProb}. Integration was stopped at $t=25,000$ (or at $t=100,000$ for simulations shown on Fig.~\ref{fig:temporal_order}). In each case the system reached a quasi-stationary state at the end of the simulations. An implementation of this particle-based system is provided in Online Resource 3.

\subsection{Order parameter}

\begin{figure}
\includegraphics[width=\columnwidth]{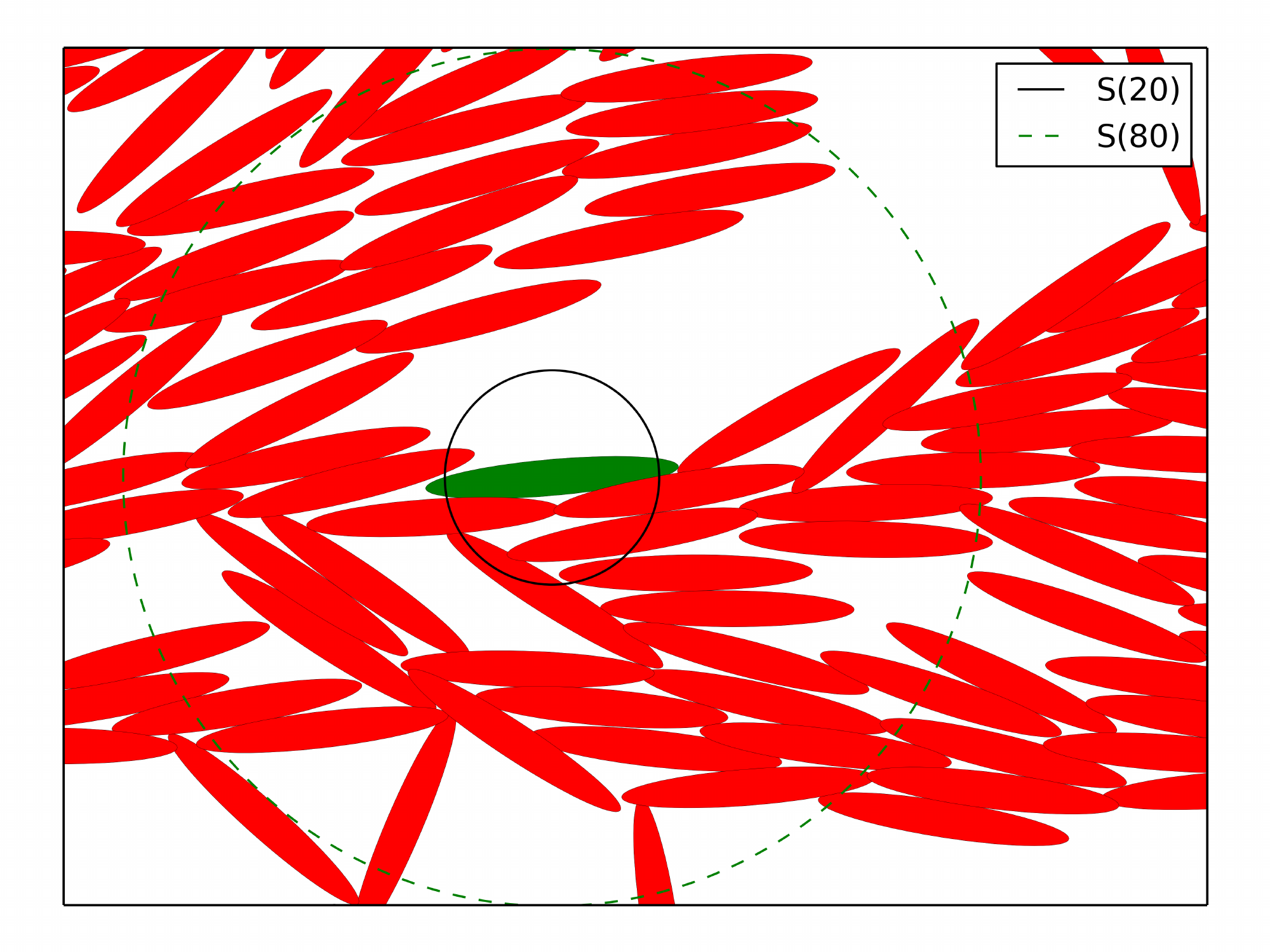}
\caption{\textbf{Ranges of local order.} Two values of $r$ for the order parameter calculation shown for the centre cell within a typical cell arrangement. While $r=20$ describes order at the local cell neighbourhood, $r=80$ is an indication of order in multicellular structures in the vicinity}
\label{fig:order_visual}
\end{figure}

Local alignment of elongated cells has been shown to play an important role in the formation of networks previously \cite{palm_vascular_2013}. To measure alignment, a local orientational order parameter is used:
\begin{equation}
S(r)=\langle\cos \left(2\theta_i(r)\right)\rangle_i.
\end{equation}
$\theta_i(r)$ is the polar angle between $\phi_i$, the orientation of cell $i$, and $\eta_i(r)=\langle \phi_j \rangle_{\left\{j: |\pos_j - \pos_i| \leq r\right\}}$, the average cell orientation within $r$ distance of cell $i$. $S(r)$ is isotopic and takes the value of $0$ for randomly oriented cells and $1$ for perfectly aligned cells. Smaller values of the order radius $r$ describe the alignment in the close vicinity of the cell ($r=20$), or in the local structures ($r=80$, Fig.~\ref{fig:order_visual}), while any global order is captured by the global order parameter, $S_g=S(r \to \infty)$, where all the cells are considered for the calculation of $\eta_i$. The software used for measuring the order parameter is provided in Online Resource 3.

\section{Results}

\subsection{Elongated cell shape induces network formation}

\begin{figure}
\centering\includegraphics[width=\columnwidth]{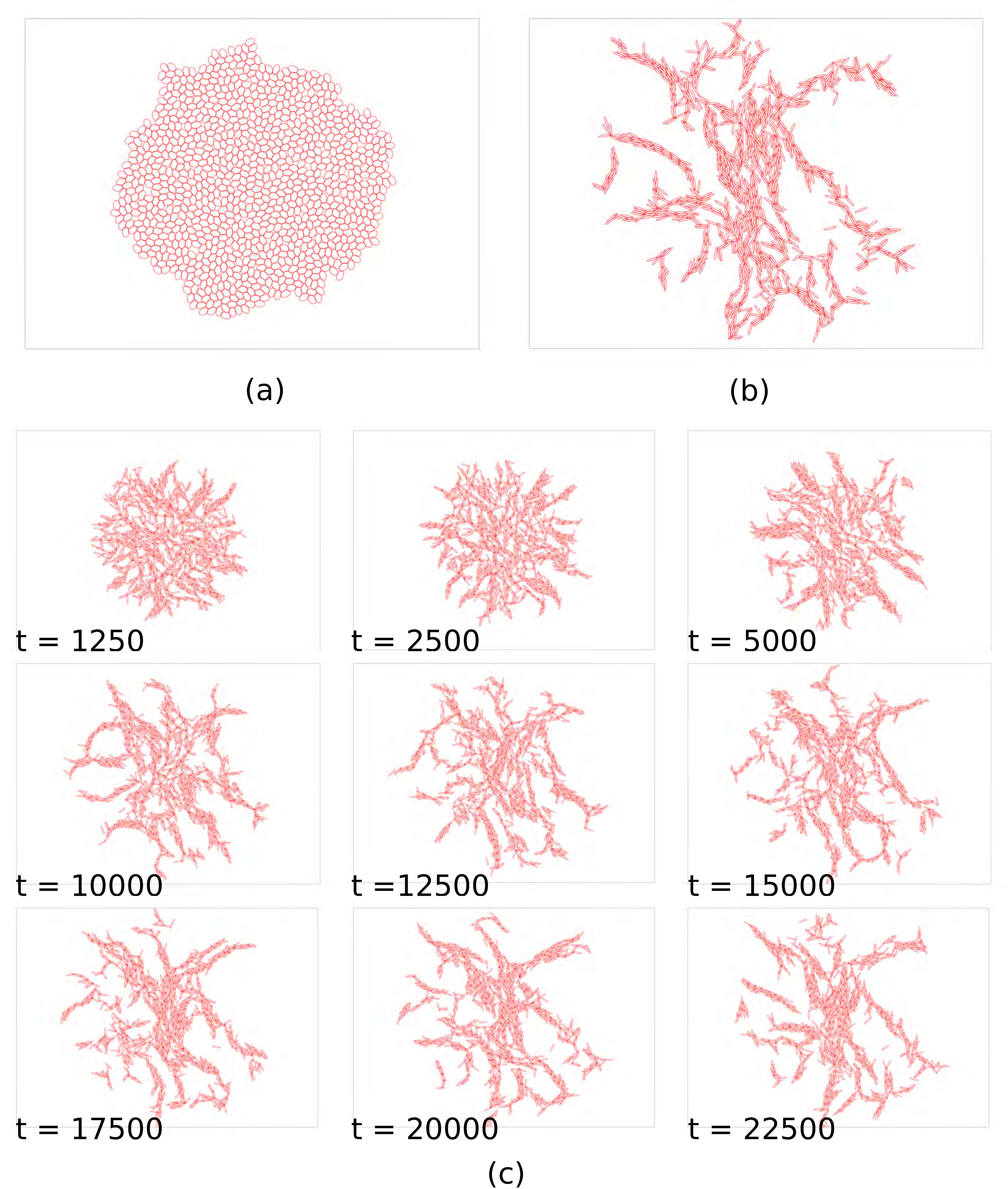}
\caption{\textbf{Cell elongation induces network formation} 
\textbf{a}, Short cells with aspect ratio $s=1.5$ form compact aggregates.
\textbf{b}, Elongated cells with aspect ratio of $s=7$ form networks.
\textbf{c}, Temporal evolution of network formation for cells with aspect ratio of $s=7$. Global order is established at about $t=10^4$. The system reaches a fluctuating steady state: the morphological details change over time, but the morphological features remain statistically constant.}
\label{fig:basic_model}
\end{figure}

\begin{figure}
\centering\includegraphics[width=\columnwidth]{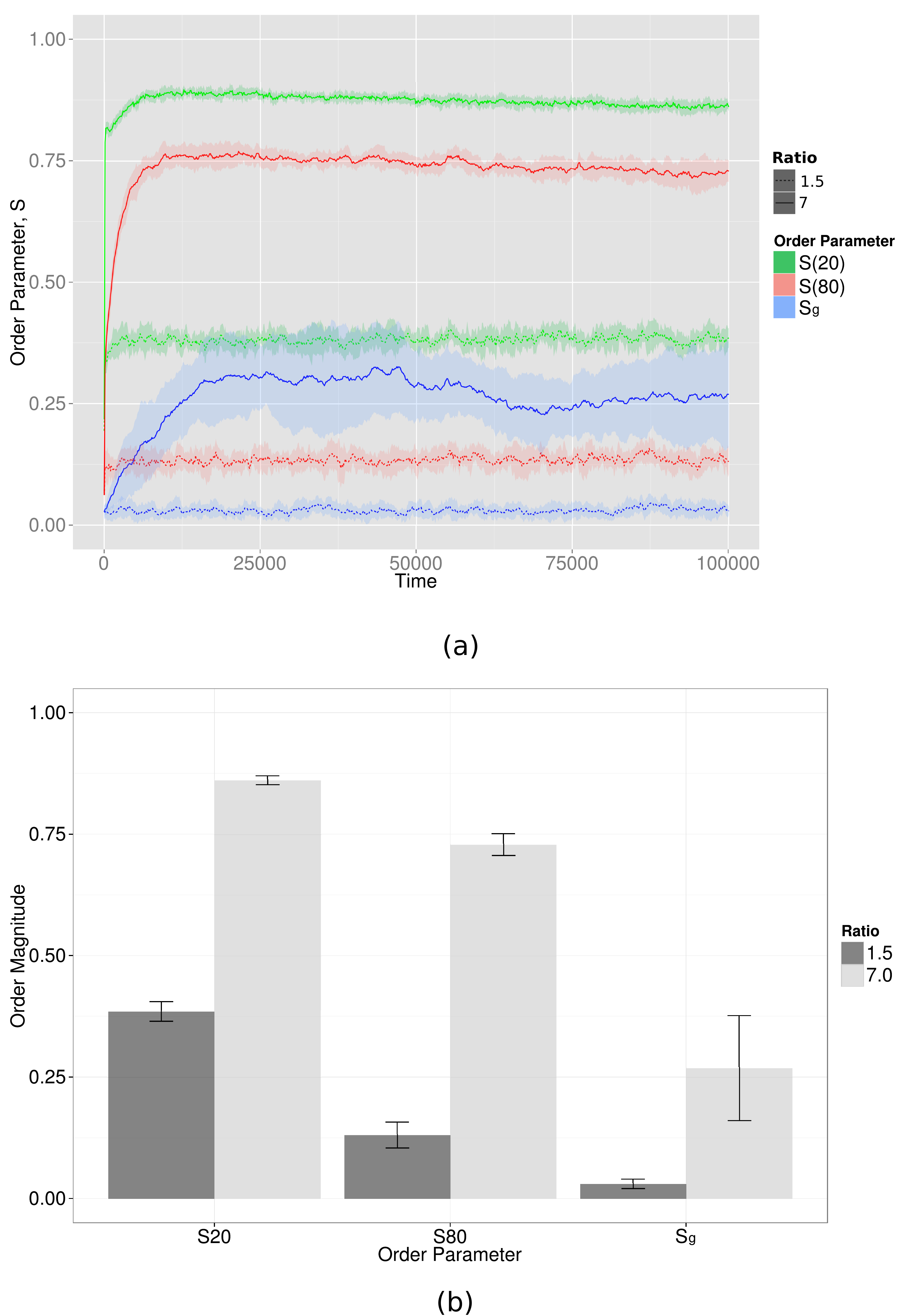}
\caption{\textbf{Order parameters of elongated and rounded cell simulations.}
\textbf{a}, Temporal evolution of the order parameter, $S(r)$, for $r=20$ (green curves), $r=80$ (red curves) and $r\rightarrow\infty$ (blue curves) for two systems of more rounded cells ($s=1.5$, dotted lines) and elongated cells ($s=7$, solid lines; $\Delta t=1$). Ribbons show standard deviations over 10 runs. Finite scale order reaches a plateau at $t=10^4$ and remains approximately stable up to at least $t=10^5$.
\textbf{b}, Order parameters from 10 independent simulations show that alignment at $t=10^5$ is higher for elongated cells than rounded cells at all length scales ($\Delta t=1$). Within configurations order at shorter scales is always higher than order at longer scales}
\label{fig:temporal_order}
\end{figure}

Rounded cells ($s=1.5$) aggregate into less ordered clusters (Fig. \ref{fig:basic_model}a and the animation in Online Resource 1), while more elongated cells with aspect ratio $s=7$ in the simulation interconnect to form elongated branches and networks (Fig. \ref{fig:basic_model}b and the animation in Online Resource 2).
The time evolution of the order parameters calculated for two simulations of round and elongated cells shown in Fig. \ref{fig:basic_model} demonstrates that the initially disordered cells align within t=$10^4$ and keep a similar level of order for at least 10 times longer at finite length scales ($S(20)$ and $S(80)$ on Fig.~\ref{fig:temporal_order}a). Positive value of the global order parameter ($S_g$) in the elongated cell configurations results from finite system size and indicates the emergence of a system-wide alignment (Fig.~\ref{fig:basic_model}c). Elongated cells ($s=7$) produce a significantly higher order than more round cells ($s=1.5$) at all length-scales (Fig.~\ref{fig:temporal_order}b). Order at the cell-to-cell range ($S(20)$) always supersedes order at the multi-cellular level ($S(80)$), which is always higher than global order ($S_g$). 

\begin{figure}
\centering\includegraphics[width=\columnwidth]{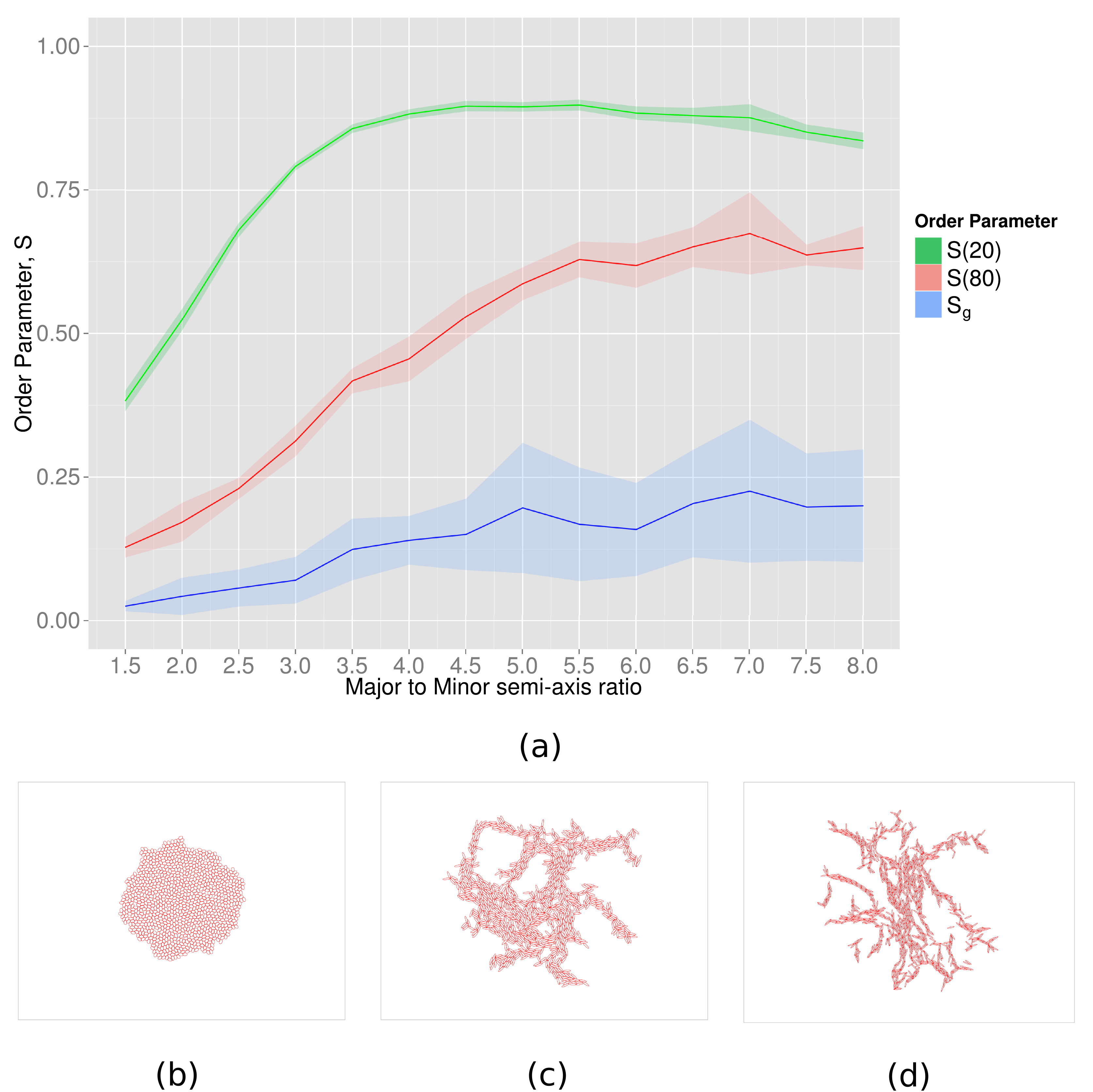}
\caption{\textbf{Emergent order in the model is an increasing function of cell elongation.} 
\textbf{a}, Simulations with uniform cells of various aspect ratios show that slightly elongated cells align locally, while more elongated cells give rise to order at longer length scales. Ribbons show standard deviations over 10 runs after $t=2.5\times10^4$.
\textbf{b--d}, Configurations at $t=2.5\times10^4$ for aspect ratios $s=1.5$, $s=4.5$, and $s=7$}
\label{fig:length}
\end{figure}

Simulations with cells of increasing aspect ratios and constant areas reveal that order is gradually increased in simulations with more elongated cells (Fig.~\ref{fig:length}). When simulations with increasing $s$ are compared, order emerges first at short length scale ($S(20)$) followed by longer range order. Global order appears in simulations with $s\geq4$. The high variation of the global order among the different simulations is consistent with the fluctuation shown on Fig.~\ref{fig:temporal_order}a. 

\subsection{Increased range of cell attraction enhances cell alignment}

Previous studies showed that chemotaxis could play an important role in vascular network formation \cite{merks_contact-inhibited_2008,merks_cell_2006}. Since chemotaxis may be interpreted in our model as an adhesive interaction over a longer range, we investigated how the range of attraction affects the observed cell alignment. 
\begin{figure}
\centering\includegraphics[width=\columnwidth]{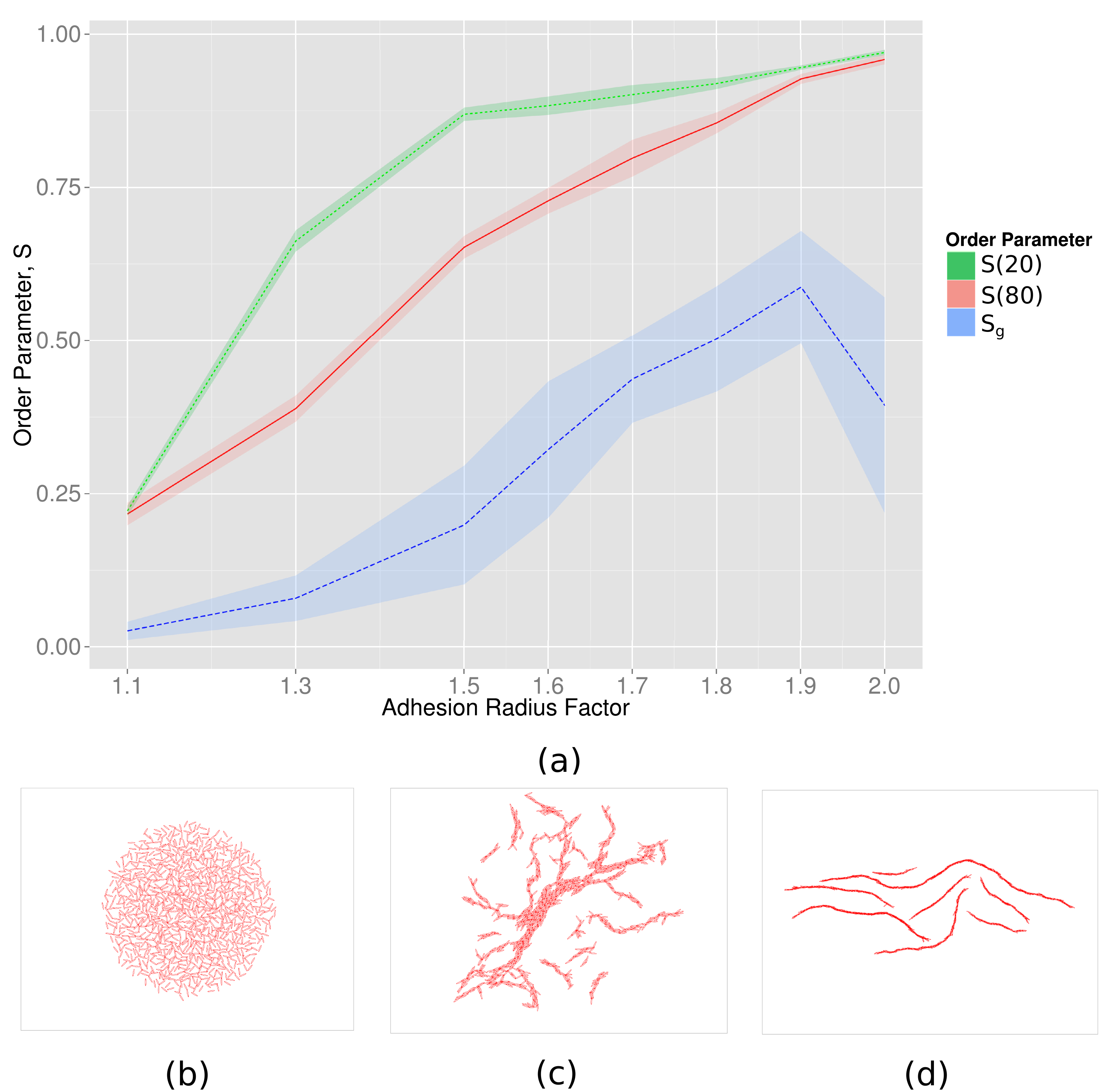}
\caption{\textbf{Increasing range of attraction increases order.} 
\textbf{a,} Local, intermediate, and global order are higher with larger attraction ranges for cells with aspect ratio of $s=7$. Ribbons show standard deviations over 10 runs after $t=2.5\times10^4$.
Configurations at $t=2.5\times10^4$ for (\textbf{b}) short,
(\textbf{c}) medium and (\textbf{d})
long attraction ranges $R_a=1.1$, $R_a=1.6$ and $R_a=2$}
\label{fig:adh_range}
\end{figure}
As expected, a short range attraction ($R_a=1.1$) results in a dissociated cell configuration and low order (Fig.~\ref{fig:adh_range}a and Fig.~\ref{fig:adh_range}b, $R_a=1.1$), compared to the control parameters (Fig.~\ref{fig:basic_model}b). A long attraction range, however, results in a marked increase in ordering, consistent with the longer range attraction of the chemotaxis studies. Interestingly, global order increases markedly at an interaction range of less than twice the cell size, where long, aligned strands of cells are formed (Fig.~\ref{fig:adh_range}d, $R_a=2$). 

\begin{figure}
\centering\includegraphics[width=\columnwidth]{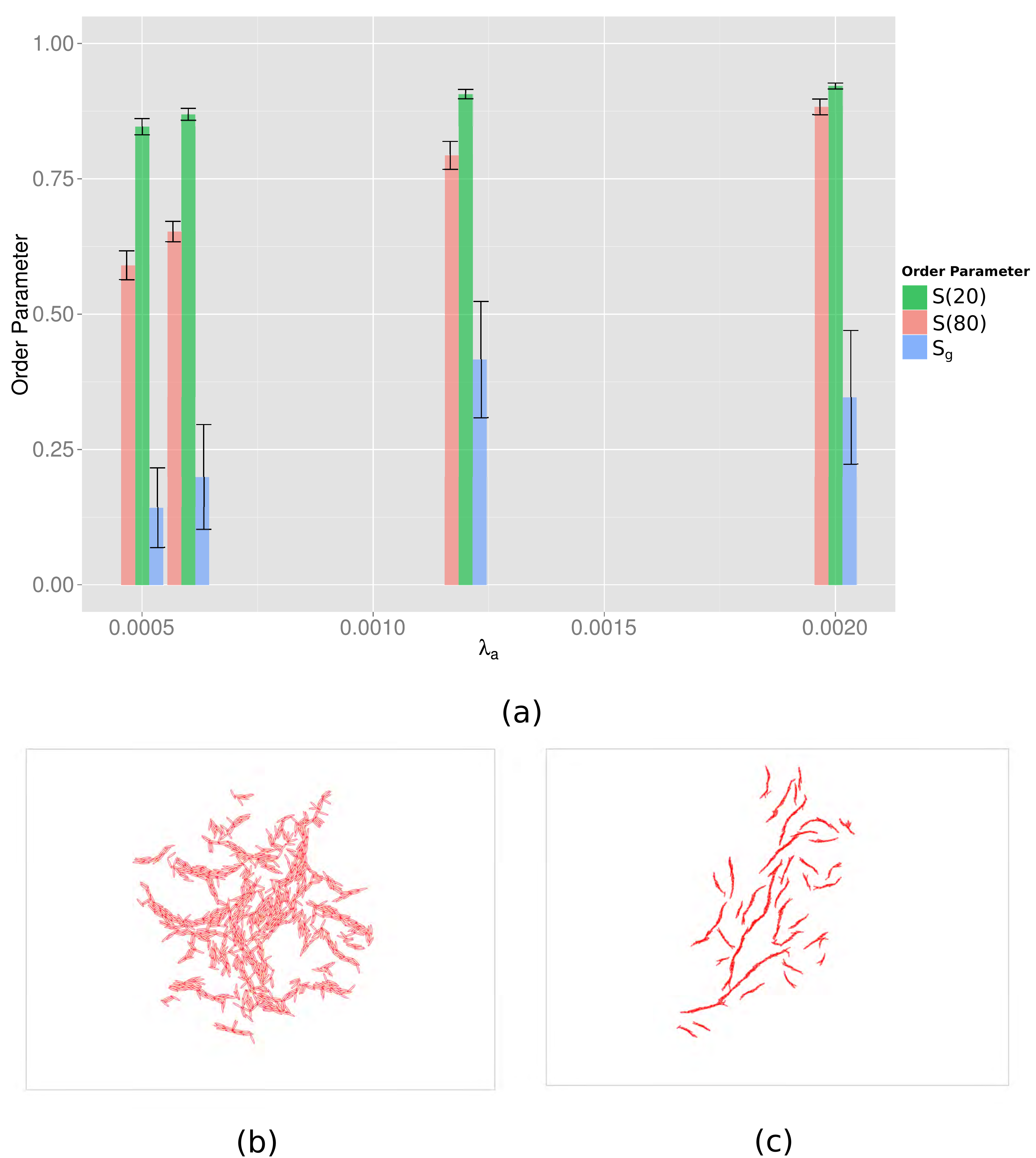}
\caption{\textbf{Stronger attraction leads to increased order.} 
\textbf{a}, Order parameters from simulations with increasing attraction parameters for cells with aspect ratio of $s=7$. 
\textbf{b}, Simulation configurations with low attraction strength $\lambda_a=4\times10^{-4}$. \textbf{c} Simulation configurations with high attraction strength $\lambda_a=2\times10^{-3}$
}
\label{fig:adh_strength}
\end{figure}

\begin{figure}
\centering\includegraphics[width=\columnwidth]{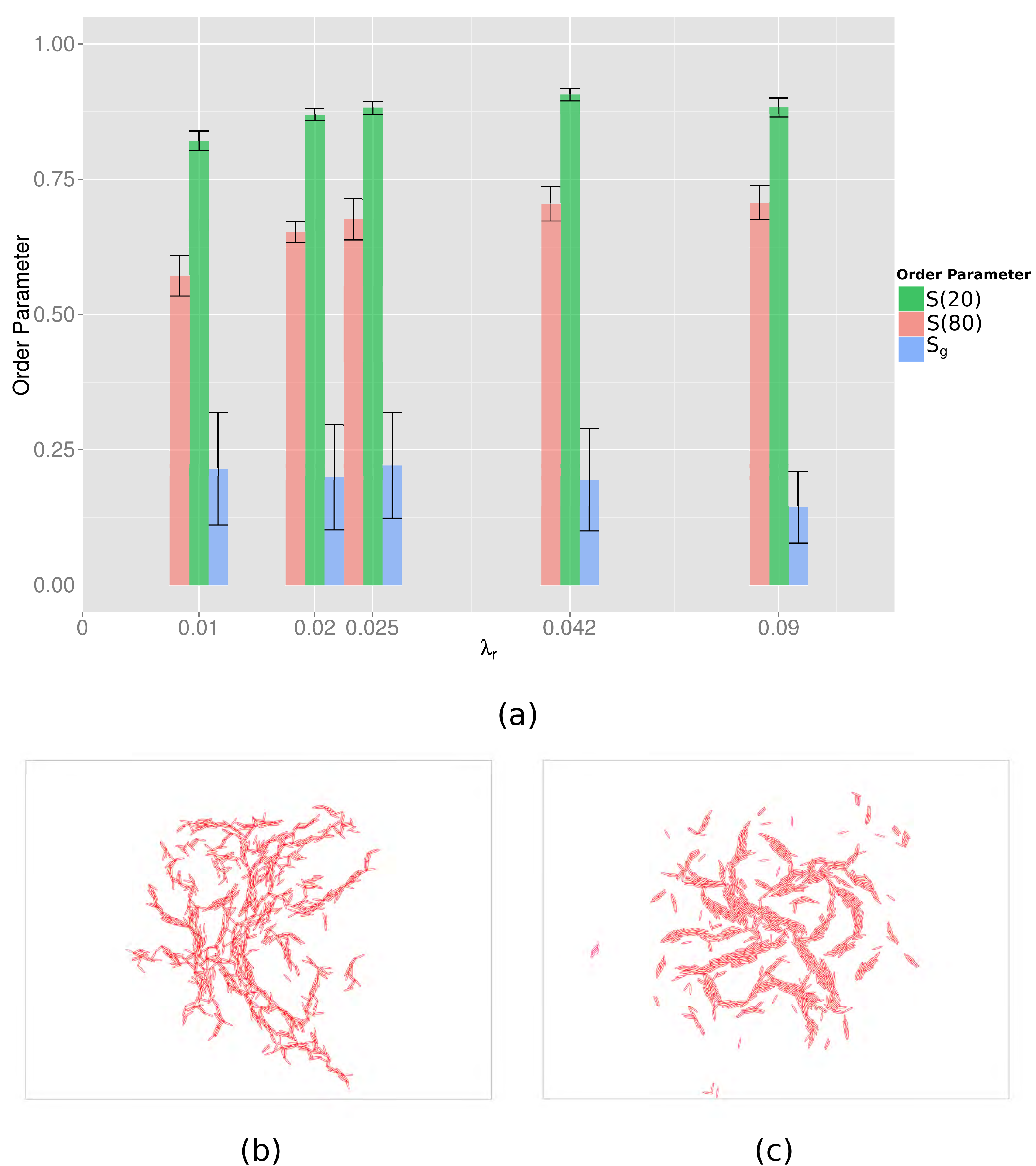}
\caption{
\textbf{Repulsion acts against order.} 
\textbf{a}, Order parameters from simulations with increasing repulsion parameters for cells with aspect ratio of $s=7$.
\textbf{b}, Simulation configurations with a low repulsion strength of $\lambda_r=10^{-2}$; \textbf{c}, Simulation configurations with a high repulsion strength of $\lambda_r=9\times10^{-2}$
}
\label{fig:rep_strength}
\end{figure}

Increased alignment is also achieved by increasing attraction strength through parameter $\lambda_a$ (Fig. \ref{fig:adh_strength}), while increased repulsion leads to dissociation and consequent loss of order (Fig. \ref{fig:rep_strength}). 

\subsection{Noise is not essential for alignment}

\begin{figure}
\centering\includegraphics[width=\columnwidth]{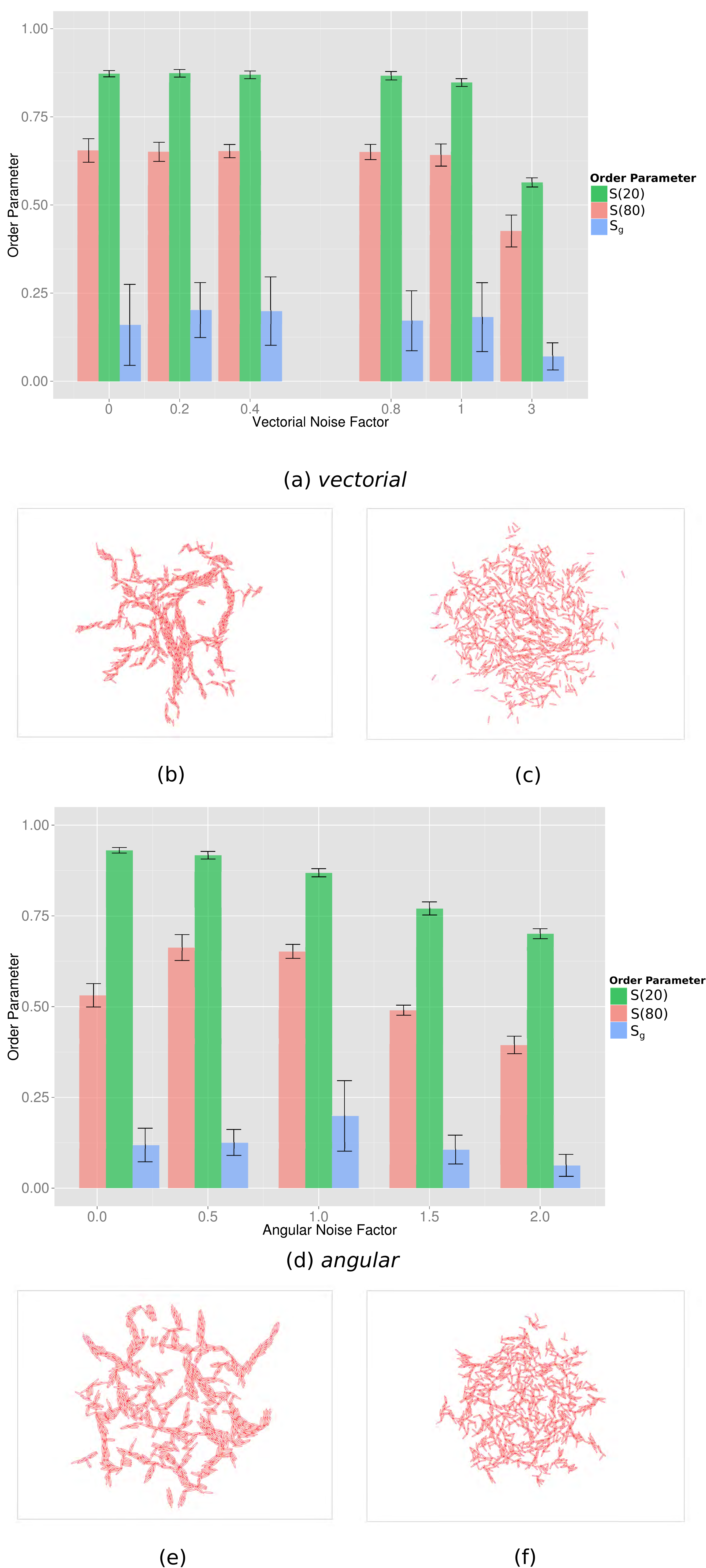}
\caption{
\textbf{Effect of noise on alignment.} 
\textbf{a}, Order parameters as a function of vectorial noise for cells with aspect ratio of $s=7$; 
\textbf{b}, Simulation configurations in absence of translational noise;  
\textbf{c}, Simulation configurations with increased translational noise ($N_v=3$); 
\textbf{d}, Order parameters as a function of angular noise for cells with aspect ratio of $s=7$; 
\textbf{e}, Simulation configurations with low angular noise ($N_a=0.1$); 
\textbf{f}, Simulation configurations with increased angular noise ($N_a=2$). Order parameters and configurations are shown at $t=2.5\times10^4$ simulation time
}
\label{fig:noise}
\end{figure}

We assessed the importance of noise for cell alignment in our model (Fig.~\ref{fig:noise}). In the absence of translational noise (Fig.~\ref{fig:noise}b) or low rotational noise (Fig.~\ref{fig:noise}e) cells align and form networks. At high noise levels the two types of noise act in similar manner; high translational noise results in a decay of global order (Figs. \ref{fig:noise}a and \ref{fig:noise}c), while high angular noise allows for more energy inefficient rotations, thus decreasing global order (Figs.~\ref{fig:noise}d and \ref{fig:noise}f). At low noise levels, the two types of noise act differently; while low translational noise has no or very little effect on the orientational order (Fig.~\ref{fig:noise}a), low angular noise results in higher local order and lower global order (Fig.~\ref{fig:noise}e), suggesting the formation of isotropic branches that point in different directions.

\subsection{Cell elongation is essential for alignment}

\begin{figure}
\centering\includegraphics[width=\columnwidth]{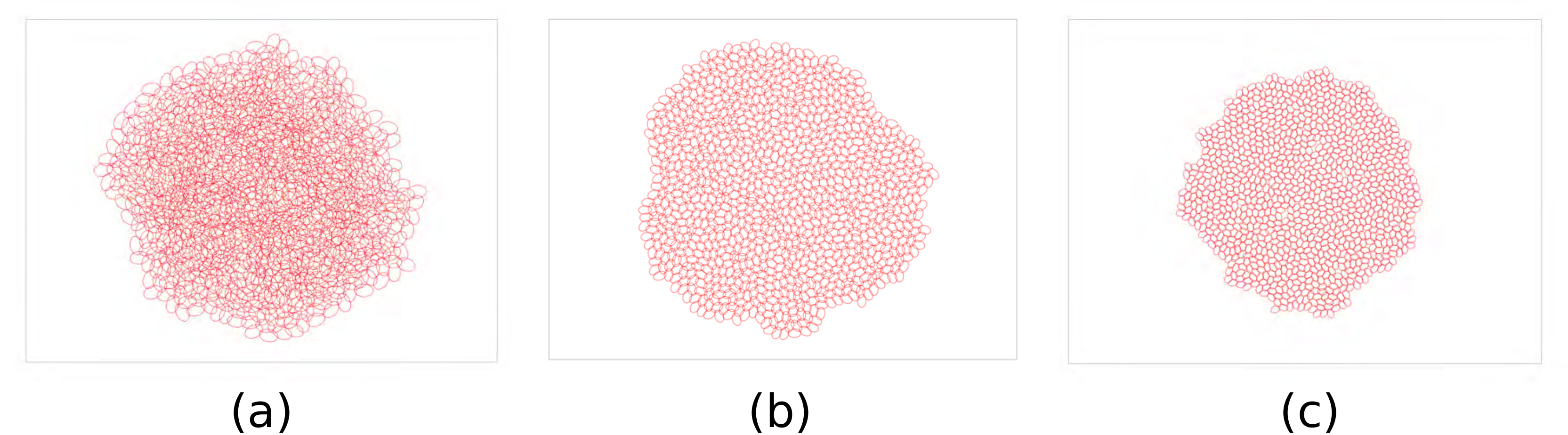}
\caption{
\textbf{Effect of parameters on alignment without cell elongation.} 
\textbf{a}, Simulation configurations with increased radius of attraction $R_a=2$;  
\textbf{b}, Simulation configurations with increased attraction strength $\lambda_a=2\times10^{-3}$; 
\textbf{c}, Simulation configurations with low repulsion strength $\lambda_r=10^{-2}$. Configurations shown at $t=2.5\times10^4$. All cells have an aspect ratio of $s=1.5$
}
\label{fig:short_test}
\end{figure}

Finally, we tested whether cell elongation is required for alignment, by influencing the parameters that had positive impact on cell order in the previous simulations. Using rounded cells with aspect ratio of $s=1.5$, three scenarios with increased range of attraction ($R_a$), increased attraction strength ($\lambda_a$) and reduced repulsion strength ($\lambda_r$) were tested. Order parameters from these simulations together with order parameters from simulations of round and elongated cells (Fig.~\ref{fig:basic_model}a-b) are summarized in Table \ref{tab:order_short}. These indicate that neither increased adhesion strength or range, nor decreased repulsion is able to order the cells to an extent comparable with elongated cell simulations. Without elongation, cell clusters are unable to deviate from their compact aggregates (Fig. \ref{fig:short_test}).
\begin{table}[h]
\centering
\caption{Order parameters for round ($s=1.5$) and elongated ($s=7$) cells, compared with cells with aspect ratio of $s=1.5$ at $t=2.5\times10^4$ tested with parameters that favour cell ordering}
\begin{tabular}{|c|c|ccc|}
\hline
Simulation & Parameter & $S(20)$ & $S(80)$ & $S(\inf)$ \\ \hline
round cells & $s=1.5$ & 0.38 & 0.13 & 0.03 \\
elongated cells & $s=7$ & \textbf{0.88} & \textbf{0.67} & \textbf{0.23} \\
high attraction range & $R_a=2$ & 0.31 & 0.11 & 0.04 \\
high attraction& $\lambda_a=2\times10^{-3}$ & 0.38 & 0.08 & 0.02 \\
low repulsion & $\lambda_r=10^{-2}$ & 0.36 & 0.12 & 0.03 \\ \hline
\end{tabular}
\label{tab:order_short}
\end{table}

\section{Discussion}

A simple particle-based model is introduced to demonstrate that elongation can indeed aid cells to aggregate into a network.  The mechanism for network formation is simple: Cell elongation, together with attraction and core repulsion, leads to local cell alignment and increase in orientational order. The fact that this prediction holds in two completely different model implementations, our particle-based model and the CPM \cite{merks_cell_2006,palm_vascular_2013}, gives confidence that the predictions are due to the explicit model assumptions, not due to unintended properties of the simulation methodology. Time evolution of order and dependence on elongation (Fig.~\ref{fig:temporal_order}b) are similar in both implementations. Increasing the attraction range of cells in the current implementation allows the study of the transition between the pure elongation hypothesis \cite{palm_vascular_2013} to the chemotaxis and elongation hypothesis \cite{merks_cell_2006}. Extended attraction range, in the form of chemotaxis, has been shown to help in the formation of regular networks with a more defined pattern size across the whole system \cite{merks_cell_2006}. Consistently with this previous result, we found that a longer range of attraction results in higher order and a more network-like structure (Fig.~\ref{fig:adh_range}). 

Cell movement dynamics is a marked difference between the CPM and our model. In the CPM cell movement emerges from the displacement of the cell boundaries, yielding  a more amoeboid cell movement. By contrast, in particle-based models model cells translate as a single unit, including translocation and rotation in our case. Cell movement in the model is described by an overdamped dynamics, where the damping factor ($\alpha$ in Eq.~\ref{eq:diff1}) controls the movement persistence of cells. For large values of $\alpha$, the left-hand-side of Eq.~\ref{eq:diff1} vanishes and Eq.~\ref{eq:diff1} can be approximated by an algebraic equation giving the cell velocities as a function of the external forces acting on the cells.  We have chosen here to work with a differential equation description, thus keeping the damping factor in our model as an explicit parameter. Although we have here only studied overdamped kinetics, smaller values of $\alpha$ could mimic persistent cell motility:  it takes time for a cell to change its direction if, e.g, the chemoattractant gradients change; the actin cytoskeleton needs to reorganize which takes time. Such persistent motion was also explicitly included in early chemotaxis-based partial-differential equation models of vascular network formation \cite{Gamba:2003dz}. Although persistence of motility was later shown to be unnecessary for network formation, it was argued that in the CPM the cell shape may cause some directional persistence (Figure~11 of Ref.~\cite{merks_cell_2006}).  In the present work directional persistence is severely reduced and is largely independent of the cell shape, suggesting that persistence is not required for network formation.

Stochasticity in our model is introduced through the \textit{angular} noise similar to the noise of the Vicsek model~\cite{vicsek_1995_phase} and the \textit{translational} noise of the Gr\'{e}goire -- Chat\'{e} model \cite{gregoire_onset_2004}. Note that these noise factors describe fluctuations in the system, and therefore the considered noise factors are independent from one another and from the noise in the past. Our results from the study of the \textit{angular} noise show that noise hinders the global order, in good agreement with previous particle-based simulations using only \textit{angular} noise and force-dipoles to describe multicellular structure formation \cite{Bischofs2006}.

Interestingly, both an increase in attraction range (Fig.~\ref{fig:adh_range})  and attraction strength (Fig.~\ref{fig:adh_strength}) leads to a higher tendency to global ordering as packing of the cells becomes tighter. This is similar to the previous observation in the CPM models \cite{merks_cell_2006,palm_vascular_2013}, where the more adhesive cells become more compact and highly ordered in domains, but without the emergence of global order. 
In the CPM cell shape is an emergent property of the cells and therefore cells are able to deform in order to maximise the contact surface and compactness (see Fig.~4a of \cite{palm_vascular_2013}). In contrast, cells in our implementation are unable to deform such that they might be less able to accommodate the boundaries of such locally ordered domains. This may force them to align globally through the compacting force of the strong attraction. Secondly, when attraction range is increased to a distance of two cell diameters in the current implementation, alignment order spans through domains of at least 1000 cells, with strands of cells aligned in parallel even without tight packing (Fig.~\ref{fig:adh_range}d, $R_a=2$). 
An effect contributing to this global ordering may be the implementation of noise in the current, particle-based implementation. In the CPM, at high densities, cells hinder each other's  translation and rotation, resulting in an increasingly slow development of the pattern as the branches grow \cite{palm_vascular_2013}. By contrast, in the particle-based model translation is not hindered by adjacent cells and occurs for cells as a whole (Eq.~\ref{eq:diff1}), while cells continue to rotate independently as a whole as long as conflicts with adjacent cells persist (Eq.~\ref{eq:rotationProb}). 

In conclusion, here we introduced a particle-based model to re-examine a hypothetical mechanism for the formation of microvascular networks: i.e., that elongated vascular cells tend to aggregate into branches of a network structure. Our previous work \cite{merks_cell_2006,palm_vascular_2013} simulated this potential mechanism using the cellular Potts model, demonstrating that indeed elongated cell shape, in combination with mutual attraction or adhesion suffices for the formation of network-like patterns. Here we have shown that this phenomenon also occurs in a lattice-free particle-based model, adding confidence that the effect is not caused by artifacts of the cellular Potts model or of the present, particle-based model. Thus  our models suggest that network formation is a natural emergent property of elongated, adhesive objects in a stochastic system. Because a range of alternative mechanisms for network formation and angiogenic sprouting have been suggested (reviewed in~\cite{merks_modeling_2009} and \cite{czirok_endothelial_2013}, see also \cite{VanOers2014}), to what extent the present mechanism contributes to angiogenesis {\em in vivo} or {\em in vitro} at this point must remain an open question. These and similar studies, however, help to generalize and categorize the main requirements for network formation and angiogenic sprouting, thus contributing to ongoing efforts to identify the controlling factors of angiogenesis from a biophysical point of view.

%
%

\bigskip
\noindent \textbf{Online Resource 1} Simulation video of rounded cells for $t=2.5\times10^4$. SImulated cells of aspect ratio $s=1.5$ form aggregates. Every frame is after $t=50$. Every second of video is $t=1250$;  see \url{https://youtu.be/BgQwd1aGxAI}

\noindent \textbf{Online Resource 2} Simulation video of elongated cells for $t=2.5\times10^4$. Simulated cells of aspect ratio $s=7$ form networks. Every frame is after $t=50$. Every second of video is $t=1250$; see \url{https://youtu.be/9zUGEBk6pio}

\noindent \textbf{Online Resource 3} C++ implementation of the presented particle-based system, and software for measuring the order parameter

\begin{acknowledgements}
Dimitrios Palachanis has completed this work during an M.Sc. research internship at CWI, as part of the Leiden University M.Sc. program Computer Science, Bioinformatics track. His internal supervisor Dr. Erwin Bakker is warmly thanked for support and guidance during the project.
\end{acknowledgements}



%
%

\end{document}